\documentstyle[preprint,prd,aps]{revtex}
\begin{document}
\draft
\preprint{
\vbox{
\halign{&##\hfil\cr
	& AS-ITP-97-15 \cr
	& hep-ph/9706443 \cr
	& May 1997 \cr}}
}
\vskip 0.5cm
\title{$J/\psi$ Electromagnetic Production at $e^+e^-$ Colliders}

\author{Chao-Hsi CHANG$^{\S,\dagger}$, Cong-Feng QIAO$^{\S}$
and Jian-Xiong WANG$^{\S,\ddagger}$}
\address{ $^{\S}$ CCAST (World Laboratory), P.O. Box 8730, 
 Beijing 100080, China.\footnote{Not mailing address for
C.-H. CHANG and J.-X. WANG.}}
\address{ $^\dagger$ Institute of Theoretical Physics, 
 Academia Sinica, Beijing 100080, China.}
\address{ $^\ddagger$ Institute of High Energy Physics,
 Academia Sinica, Beijing 100039, China.}

\maketitle

\begin{abstract}

Within the framework of QED we investigate the inclusive $J/\psi$ production
at $e^+e^-$ colliders.  It is expected to be as a further
test to the charmonium production mechanisms and QED. 
We find that at the energies of CESR, BEPC, and TRISTAN
(e.g. $4.0 GeV \leq \sqrt s \leq 60 GeV$), 
the contributions of the concerned electromagnetic processes 
to the $J/\psi$ inclusive production are great, even 
dominant, thus they greatly affect the observation 
of the color-octet $J/\psi$ production signature at
$e^+e^-$ colliders. The production of
$\psi'$, being similar to that of $J/\psi$'s, is roughly estimated,
and its influence on the observation of the color-octet 
$J/\psi$ production signature in $e^+e^-$ collision is also
discussed.


\end{abstract} 

\vfill \eject


Since $J/\psi$ was discovered, the issues on its production has constantly
drawn a lot of attention from both theorists and experimentalists,
especially after several important progresses 
being made recently. The first is about
the calculation of the fragmentation
functions of a parton evolving into 
a heavy meson (heavy quarkonia or $B_c$ meson) in 
the framework of perturbative QCD\cite{frag}.
The second comes from the comparison of the theoretical predictions,
which are achieved due to the calculations of the fragmentation
functions, with the experimental data of the $J/\psi$ and 
the $\psi'$ prompt production\cite{data}.
As a result of the comparison,
a disparity between the theoretical predictions and 
the Tevatron experimental data, the so-called 
`$\psi' (\psi)$ surplus' problem, is found\cite{plus}. The third
advancement is the advent of a plausible and interesting 
suggestion, the so-called `color-octet mechanism' in quarkonium production
\cite{octet}, in solving this puzzle.  To confirm the color-octet
mechanism further, several suggestions have been proposed.
Many a kind of production processes
of the $J/\psi$ production in colliders are investigated by groups
\cite{flem,ee,photon,cho1,octets,fixed,braaten,cho}.
Among them the $J/\psi$ inclusive production
in $e^+e^-$ annihilation has received special
and broad interests \cite{braaten}. 
Being one of the best grounds to observe the color-octet signature
because of the clean background, 
$e^+e^-$ collider facilities are
emphasized by the authors\cite{braaten,cho,ktchao}. 
Whereas, we find that there are two kinds of $J/\psi$ production processes 
in $e^+e^-$ collision, i.e., i).
the scattering production $e^+ + e^-\to e^+ + e^- +J/\psi$ 
(we will refer this kind of $J/\psi$ production as scattering production
later on in this paper)
depicted as the Feynman diagrams Figs.1.a, 1.b.
ii). the hard-photon-associated production 
$e^+ + e^-\to J/\psi +\gamma$ 
(we will refer this kind of $J/\psi$ production as hard-photon production
later on in this paper)
depicted as Feynman diagrams Fig.1.c, are overlooked in literatures, which
have certain degrees of influence on the conclusions of the previous work.
Recently the hard-photon production $e^+e^-\to J/\psi +\gamma$ in the $e^+e^-$ 
annihilation was computed, and we found in that paper that this kind of
process indeed has a great influence on the observation of the color-octet 
signature at $e^+ e^-$ colliders \cite{cchw}. Furthermore, we believe it is 
meaningful to consider these two kinds of $J/\psi$ production processes
(scattering and hard-photon production)  together in detail to give a more
complete estimation about their effects on the observation of the
color-octet signature. Although there are still other kinds of $J/\psi$
production processes left unconsidered in $e^+ e^-$ annihilation, 
however, they are expected to be neither as great  nor as important as the
scattering one in affecting the observation of the color-octet signature. 
e.g., $e^+ +e^-\to J/\psi +f +\bar f$,  where $f, \bar f$ denote a pair of
quark and  anti-quark or lepton and  anti-lepton except for a pair of
electron and positron, is similar to the scattering production process
$e^+ +e^-\to e^+ +e^- +J/\psi$ at the first glance, whereas in fact they
are much different. The former, being always through $e^+ e^-$ s-channel
annihilation,  suffers from the s-channel suppression due to the
annihilation, especially when the collision energy increased to a certain
scale. Therefore, We will not discuss these kinds of production here,
but leave them to a later analysis elsewhere\cite{cchw1}. 

Theoretically, from Figs.1.a-1.c it is easy to see that the corresponding
processes are of QED nature. A common features of them  is that 
there is always a virtual photon coupling to an electron in relevant
diagrams, which stands as the photon in the $J/\psi$ resonance.
Thus, except for the non-perturbative QCD factor, the processes are of the
pure QED. It is this reason that we call this two kinds of $J/\psi$
production as the electromagnetic production. And because of their this
kind of character they may be computed reliably and precisely
to any given order in $\alpha$. Note that for simplicity, only 
the typical Feynman diagrams for the $J/\psi$  electromagnetic
production is presented here.  Of the scattering production processes of
$e^+e^-\to e^+e^- +J/\psi$ depicted as Fig.1.a and Fig.1.b, the later diagram
(and the other similar ones which are not shown) contains a exchanged photon 
and an electron (or positron) in t-channel, thus the diagram is expected
to contribute a dominant fraction to the cross section. Especially,
in the area of the phase-space where the momenta of the  exchanged photon
and the electron approach to their mass shells, respectively. 
The hard-photon production, with the character of a hard photon emitting
from the electron line and the $J/\psi$ coupling to the electron line
through a virtual photon as well, is depicted as Fig.1.c. It is of order
$O(\alpha^3)$ in QED coupling constant, i.e., one order lower  
than that of scattering ones. Besides,  because there is
an exchanged electron (or positron) in t-channel as well, it will
make a great contribution in the $J/\psi$ production.

Because of the concerned energy scales here are comparatively
rather `low', which may even comparable to $m_{J/\psi}$, the so-called
electromagnetic fragmentation approach (EMFA)\cite{flem} is not
applicable. The EMFA works well at much high energies ($\sqrt s \gg
m_{J/\psi}$). In addition, if one adopts the electromagnetic fragmentation
approach here regardless of whether it is suitable or not, it seems to be
no advantages in simplifing the calculation of the processes.
Therefore, all the results in this paper come from a full QED calculation 
rather than EMFA approach. 

The electromagnetic  production processes emphasized here are all in
color-singlet production mechanism, therefore, at various energies of
present available collider facilities, such as TRISTAN, CESR and BEPC, to
calculate the processes precisely is meaningful to the study of the
color-octet production mechanism. On the other hand, the estimation of the
$J/\psi$ electromagnetic production may play a role in further testing
the applicability of the QCD in lower energy region like $m_{J/\psi}$ or
so. Hence, to make a thoroughly discussion on the $J/\psi$ electromagnetic
production is needed.

The $J/\psi$ production cross section of the hard-photon  process can be
calculated straightforward making use of the standard formalism \cite{p8}.
The differential cross section can be expresses as:
\begin{equation}
\displaystyle \frac{d\sigma}{dt}=\frac{32\pi\alpha^3|R_S(0)|^2}
{3M^3s^2} [\frac{2M^2 s}{tu}+\frac{t}{u}+\frac{u}{t} ],\\[2mm]
\end{equation}
where $M$ is the mass of charmonium; $\alpha$ is electromagnetic
coupling constant; $s=(p_1+p_2)^2; \;\; t=(k-p_1)^2; \;\; u=(P-p_1)^2.$
For simplicity, the eq.(1) is a result of throwing away the electron mass, 
while in doing numerical calculations the mass will be kept. As the radial
wave function at the origin $|R_S(0)|^2$ here is exactly equal to that
appearing in the corresponding equation of $J/\psi \to e^+e^-$, i.e.
\begin{equation}
\displaystyle \Gamma(J/\psi \to e^+e^-) = \frac{4\alpha^2}{M^2}
|R_S(0)|^2, \\[2mm]
\end{equation}
we will determine it from the  experimental value of the 
decay width of $J/\psi$ to lepton pair. This procedure provides all of the
theoretical corrections to the wave function at origin being included.

The differential cross section versus $t$, hence the angular $\cos\theta$,
at a given CMS energy is
\begin{equation} 
\displaystyle \frac{d\sigma}{dt}=
\frac{6\pi\alpha\Gamma^{exp.}(J/\psi\to e^+e^-)} {M s^2} 
[\frac{2M^2 s}{tu}+\frac{t}{u}+\frac{u}{t} ]. \\[2mm]
\end{equation}
Therefore,
\begin{equation} 
\displaystyle \frac{d\sigma}{d\cos\theta}=
\frac{6\pi\alpha\Gamma^{exp.}(J/\psi\to e^+e^-)s} {M(1-r)\sin^2\theta} 
[(1+r)^2+(1-r)^2\cos^2\theta], \\[2mm]
\end{equation}
where $r\equiv M^2/s$.

The analytical formulae of the cross section of scattering production
processes are complicated and hence are not suitable to list here. In
fact, in doing the numerical calculations, we have used our computer
programme code. And to test its reliability we have compared some results
of using programme code with that using the full analytical formulae.
The value of the wave function at the origin used in these processes 
take the same as that in the hard-photon production.

The total cross sections for various kinds of the $J/\psi$ production
processes in $e^+e^-$ collision are plotted in Fig.2. As pointed out
above, for we are also interested in seeing the influence of the 
$J/\psi$ electromagnetic production on the observation of the color octet
signature through detecting the inclusive $J/\psi$ production, 
in Fig.2  we show a comparison of the contributions of
the $J/\psi$ electromagnetic production with that of 
the color-octet ones depicted as Fig.1.d and the color-singlet one as
Fig.1.e, which have been studied and stresses in Refs. \cite{cho,ktchao,cho1}.
In this paper, the corresponding results are obtained directly from those
references.

As discussed above, of the scattering production, the relevant Feynman
diagrams may be divided into two sub-groups: one always contains an
$e^+e^-$ annihilation (Fig.1.a) and the other always contains a
t-channel exchange of a photon (Fig.1.b). For the processes containing a
exchanged photon in t-channel we will call it as the `scattering group' in
this paper. It is easy to check that each group itself is gauge invariant. 
To see the different contributions from each group, in Fig.3 we plot the
cross section not only for each sub-group, but also for the total as well. 
From Fig.3 one can see obviously that the scattering sub-group's 
contribution to $J/\psi$ production is dominant over the other. With
the same reason, the scattering production will be the greatist one
among various kinds of production processes at high energies, although it
is at least one order higher in $\alpha$ than most of the others.
The similar production process $e^+e^-\to J/\psi + f\bar f$ ($f \neq e$),
as discussed above, containing the Feynman diagrams of the type of Fig.1.a
only, is expected to have a contribution minor to that of the scattering
one's to the inclusive $J/\psi$ production in $e^+e^-$ annihilation.

The differential cross sections of $d\sigma/d\cos \theta$ of different
processes as shown in Fig.1 are plotted in Figs.4.a-4.c, where the
$\theta$ denotes the angle between the direction of the produced $J/\psi$
with the colliding beam. The CMS energies are taken the ones of the
colliders BEPC, CESR and TRISTAN, respectively. In order to see the
dependent behavior of the differential cross section  of the $J/\psi$
production on the energy more precisely, the curves for the differential
cross sections at several energies are plotted in Fig.5.

Actually, a complete set of the Feynman diagrams should contains
$Z$-boson-exchange ones in replacing of the virtual photon in each
diagram of Figs.1.a -1.e. Whereas at low energies, such as at BEPC, CESR,
even TRISTON, where $\sqrt s < m_Z$, all the contributions from the 
$Z$-boson-exchange diagrams are small, thus, can be neglected. At high
energies, e.g., equal or higher than that of LEP-I, the propagator of the
virtual $Z$ boson may approach or overtake its mass pole, while the
contribution from the $Z$ boson exchange may becomes great. Therefore we
cannot always overlook them  without a careful study. To show the virtual
$Z$ effect, we have included this kind of contribution in our numerical
calculations, which can be seen as a peak in Fig.3 around the energy of
$Z$ resonance. In general at such a high energy, many complicated $J/\psi$
production channels may be open, therefore, more detailed investigations
are still needed. At present, we only restrain ourselves to a relatively
low energy situation. 

From Fig.2, one can see that the contribution from the scattering
production at comparatively high energies and that from the hard-photon
production at comparatively low energies are dominant among all kinds of 
the considered production processes. Furthermore from Figs.4.a-4.c one
may see the fact that the $J/\psi$ electromagnetic production
has the common features that the differential cross section of the
production approaches to the maximum when the produced $J/\psi$
approaches to the beam direction, but it is still significant when the
produced $J/\psi$ at large $P_T$. This is due to the fact that both cases
are dominated by t-channel exchange diagrams. The produced $J/\psi$ in
the hard-photon and the scattering processes, the former at comparatively
low energies and the later at comparatively high energies,
in the direction perpendicular to that of the beams still contributes 
such a fraction that is not much smaller than the greatest one of  the
considered `others', no matter the greatest one of the considered `others'
is alternated with the change of the CMS energy.  For instance, the
contribution of the color-singlet one corresponding to the Feynman diagram
Fig.1.e is smaller than those of the color-octet ones corresponding to
the Feynman diagram Fig.1.d in the energy region $\sqrt s \leq 12 GeV$,
while the situation will be converse in the energy region $\sqrt s \geq
12 GeV$. Moreover, it should be noted that the hard-photon and the
scattering
$J/\psi$ production have a similar angular distribution in shape as the 
color-octet $J/\psi$ production, whereas, the shape is emphasized as
the peculiar character of the produced $J/\psi$ in color-octet
processes in detecting the color-octet signature \cite{braaten}.

In showing the energy dependence, the total cross sections versus
the CMS energies for various kinds of production processes, the Fig.2
is depicted. Considering
the fact that it is impossible to measure either the hard photon or the
pair of $e^+e^-$ when they approach to beam direction, we plot two
curves in Fig.2 for each of the scattering process. One with a cut on
the outgoing angular of $J/\psi$ and the other without any cut. From
Fig.2 one may see obviously the difference of the one without a cut with
the one  with a cut of $20^0\leq \theta \leq 160^0$  on the angular
between the direction of beam and that of the produced $J/\psi$. 

For $\psi'$ production, each of the processes having contributions to the
$J/\psi$ production will do for $\psi'$. The $\psi'$, being a 
radial excited state of $J/\psi$,  has the same quantum numbers as that
of $J/\psi$ in the non-relativistic limit. Moreover, because the
$J/\psi$ electromagnetic production has a common features, i.e. the
$J/\psi$ always couples to a charged fermion line through a virtual
photon, we must have a very similar result for the 
$\psi'$ electromagnetic production  with the same types of the Feynman
diagrams as Figs.1.a-1.c.
The difference of the electromagnetic production of these two mesons
exist only in the values of wave function at origin and the slightly
different masses. It is well known that the squared absolute values of the
wave function at the original of $J/\psi$ and $\psi'$ are different
roughly by a factor 2. Therefore, the rate of the 
$\psi'$  electromagnetic production is roughly equal to one half of that
of the $J/\psi$  electromagnetic production under non-relativistic limit.
As the branching ratio
of the decay $\psi' \to J/\psi +\cdots$ is quite great, $\sim 57\%$, in
$e^+e^-$ collision the signal of the $\psi'$ production with a prompt
cascade decay to $J/\psi$ may generate an event pattern, which looks like
as that of the direct color-octet $J/\psi$ production or of the direct
color-singlet $J/\psi$ production very much. Thus, the $\psi'$ production
and then decay into $J/\psi$ will also disturb the detection to the
color-octet $J/\psi$
production signature in a certain degree.

In our calcualtions, the input parameter values are \cite{cho1,cho,data1}
\begin{eqnarray}
\alpha_s(2m_c)=0.28,\;\alpha_{EM}(2m_c)=0.0075,\;m_c=1.48~GeV,\;
m_e=0.51~MeV,~~~
\nonumber\\
\nonumber\\
\Gamma_{ee}=5.26\pm 0.37~keV,\; <0|{\cal
O}_1^{J/\psi}(^3S_1)|0>=1.2~GeV^3,~~~~~~~~~~~~~
\nonumber\\
\nonumber\\
\frac{<0|{\cal O}_8^{J/\psi} (^1S_0)|0>}{3}+\frac{<0|{\cal O}_8^{J/\psi}
(^3P_0)|0>}{m_c^2}=(2.2\pm 0.5) \times 10^{-2} GeV^3,~~~~~~
\end{eqnarray}
which are widely used in literatures. However, it should also be noted
here that the input values of the color-octet matrix elements come from
the fitting procedure of the Non-relativistic QCD (NRQCD) \cite{nrqcd}
calculation \cite{cho1} with the Tevatron data\cite{data}. The obtained 
value there are not fully consistent with other determinations by
fitting to different experimental data, and it appears to be
overestimated. Moreover, in
calculating the color-octet $J/\psi$ production processes \cite{braaten},
both the values of $<0|{\cal O}_8^{J/\psi} (^1S_0)|0>$ and
$<0|{\cal O}_8^{J/\psi}(^3P_0)|0>$
have taken the maximum, therefore, the estimations
for the color-octet production are probably also overestimated.

In conclusion, the $J/\psi$ electromagnetic production, the hard-photon
production and the scattering production, in $e^+e^-$ collision is an
interesting and important issue, not only because it has been less
considered so far, but also because it possesses distinctive
significance to the study of quarkonium physics. It has a different
character with the ones depicted as Feynman diagrams Figs.1.d and
Figs.1.e. The electromagnetic production is dominantly through the
t-channel exchange, but the `others' are through s-channels. In addtion,
what we would like to emphasized  here is that the hard-photon production
may be used in experiment to test the QCD calculations and the quarkonium
production mechanism, if experimentally the hard photon can be identified
well, i.e. the process can be measured exclusively. Even, this process may
be uses in calibrating the detector for the relevant experiments, because
$\alpha$ is smaller enough to provide a reliable perturbative calculation.
Moreover, if
experimentally the final $e^+e^-$ pair (not from the $J/\psi$ decay) in
the process
$e^+ +e^-\to e^+ +e^- +J/\psi$ may be observed exclusively, the process
may also be used to play the same role as what hard-photon process do. If
one would like, as suggested by \cite{braaten}, to observe the signature
of the color-octet mechanism in $e^+e^-$ annihilation merely through the
inclusive $J/\psi$ production at comparatively low energies facilities, 
such as those of BEPC, CESR and TRIESTON, one has to
take into account the contribution of the electromagnetic production of
not only the $J/\psi$ but also $\psi'$. One should take that part of
events as the background of the color-octet signature's in the $J/\psi$
inclusive production. In principle, this may be practicable. In practice,
it needs a precisely exclusive measurement of the $J/\psi$ production in
order to distinguish the color-octet signature from the background
presented here. In all, there should be more investigations on the
quarkonium production, even in the electron-positron collision.
 
\vspace{1cm}
\centerline{\Large \bf Acknowledgements} 
This work was supported in part by 
the National Science Foundation of China, the Grant LWTZ-1298
of the Chinese Academy of Sciences, and the Hua Run Postdoctoral Science
Foundation of China.

\newpage
\vfill\eject

\centerline{\bf \large FIGURE CAPTIONS }

\noindent
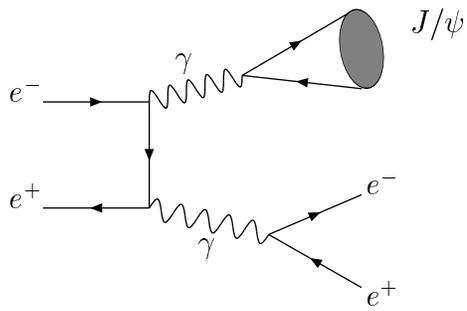
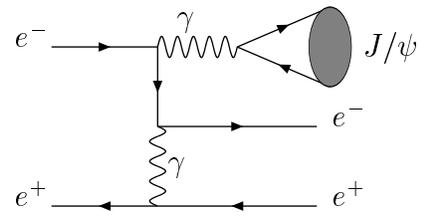
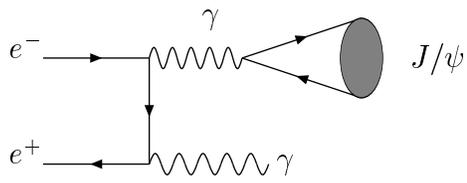
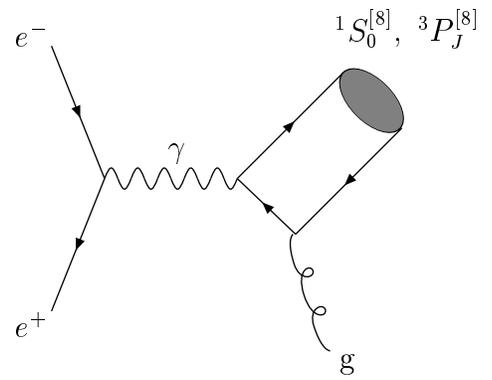
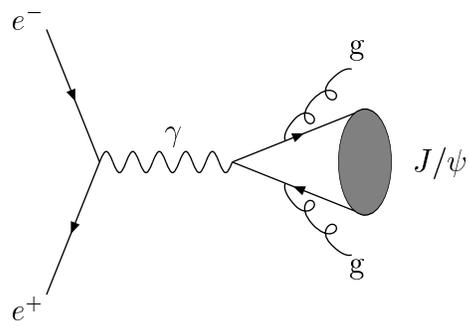
{\bf Fig.1:} The typical Feynman diagrams of the concerned $J/\psi$
production color-singlet and color-octet processes in electron-positron
collision. a) One-t-channel scattering process of $e^+e^-\to e^+ +e^-
+J/\psi$; b) Two-t-channel scattering process of $e^+e^-\to e^+e^- +
J/\psi$; c) The hard-photon process of $e^+e^-\to \gamma +J/\psi$ 
with an electron exchange in t-channel;
d) The color-octet $J/\psi$ production processes of $e^+e^-\to g +J/\psi$;
e) The traditional color-singlet $J/\psi$ production process of $e^+e^-\to
g+ g +J/\psi$.
\vskip 0.5cm
\noindent
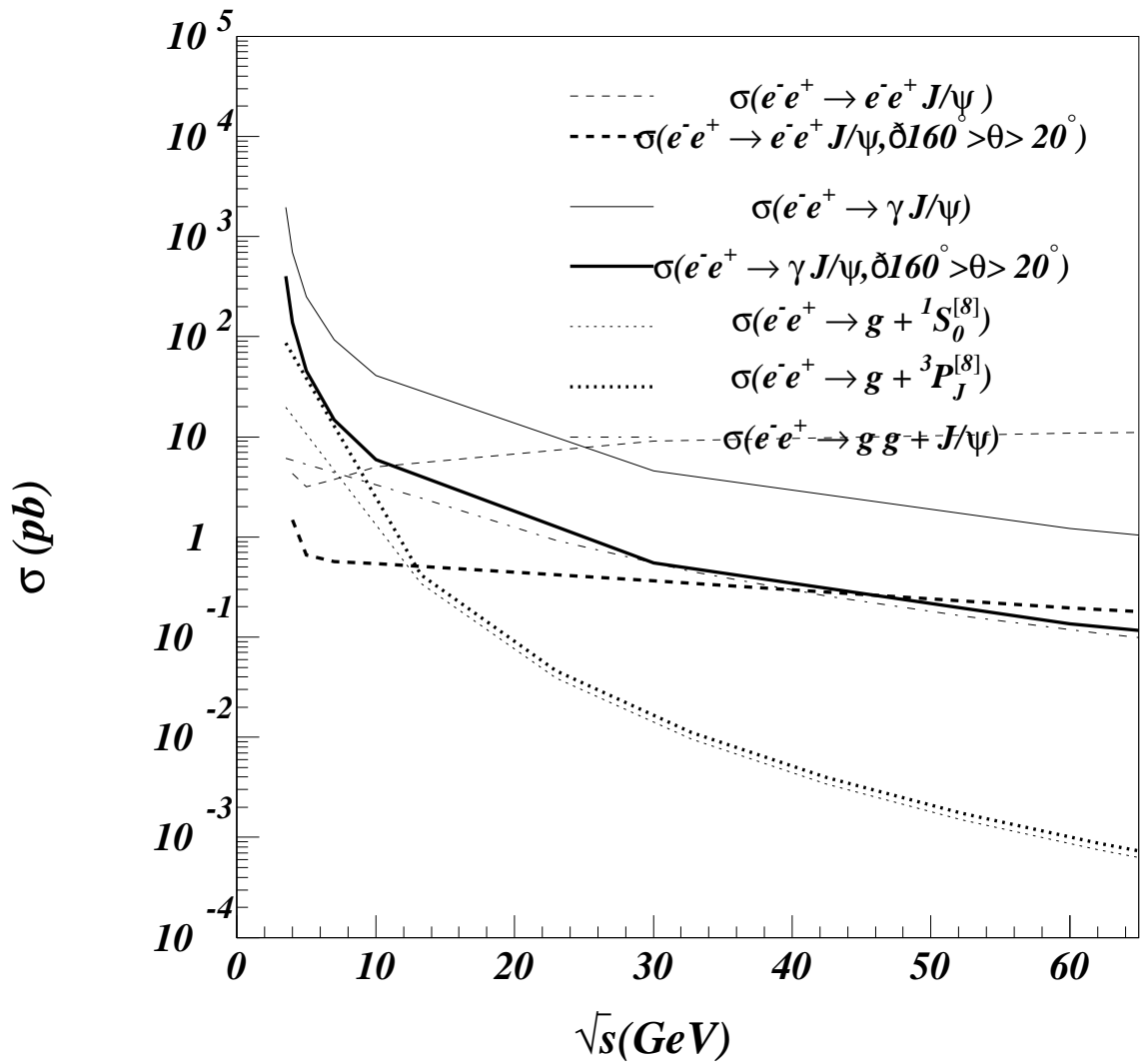
{\bf Fig.2.} The total cross sections of the $J/\psi$ production versus
the CMS energy $\sqrt s$ of various processes.
\vskip 0.5cm
\noindent
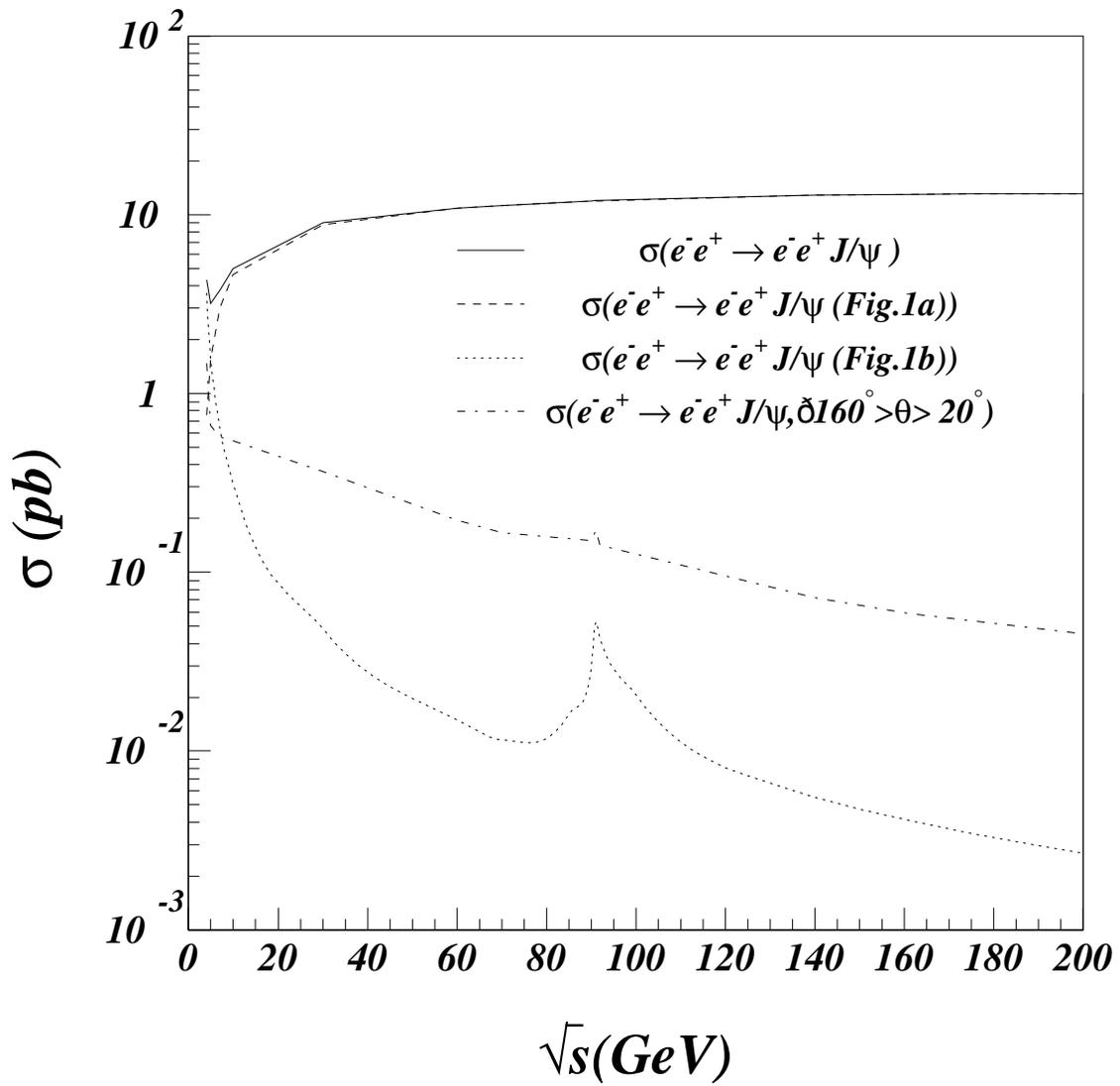
{\bf Fig.3} The total cross sections for the $J/\psi$ production
process $e^+e^- \to e^+ +e^- +J/\psi$. The thin solid line 
illustrates the summed cross section; the dashed-dotted line depicts that
with a cut $20^0\leq \theta \leq 160^0$ in angle; the dashed line denotes
the contribution only from the Feynman diagrams Fig.1a; the dotted line
denotes the contribution from Feynman diagrams Fig.1b.
\vskip 0.5cm
\noindent
{\bf Fig.4:} 

\noindent
Fig.4.a. The differential cross sections $d\sigma/d\cos \theta$ of the
$J/\psi$ production for the various processes at the CMS energy
$\sqrt s=4.03 GeV$ (BEPC). 

\noindent
Fig.4.b. The differential cross sections $d\sigma/d\cos \theta$ of the
$J/\psi$ production for the various processes at the CMS energy
$\sqrt s=10.6 GeV$ (CESR). 

\noindent
Fig.4.c. The differential cross sections $d\sigma/d\cos \theta$ of the
$J/\psi$ production for the various processes at the CMS energy
$\sqrt s=64.0 GeV$ (TRISTAN). 
\vskip 0.5cm
\noindent
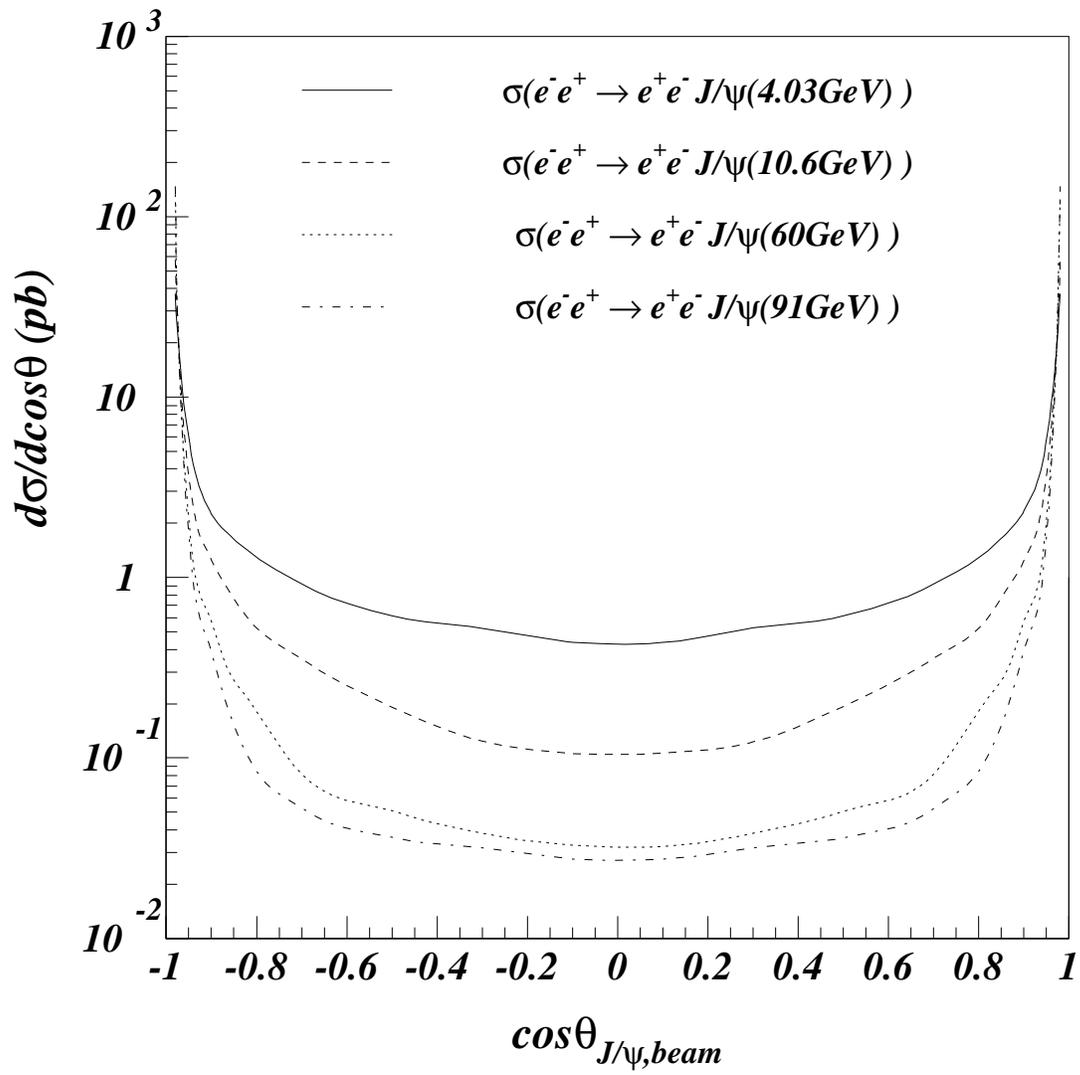
{\bf Fig.5} The differential cross sections $d\sigma/d\cos \theta$
of the $J/\psi$ electromagnetic production at different CMS energy.

\end{document}